\documentclass[journal,draftcls,onecolumn,12pt,twoside]{IEEEtranTCOM}

\normalsize

\ifCLASSINFOpdf
\else
\fi

\usepackage{amsthm}
\usepackage{amsfonts,amssymb,latexsym,cite}
\usepackage[cmex10]{amsmath}
\usepackage{algorithmic}
\usepackage{array}
\usepackage{bbm}
\usepackage{mathrsfs}
\usepackage{multirow}
\usepackage{verbatim}
\usepackage{color,xcolor}
\usepackage{graphicx}
\usepackage{epstopdf}
\usepackage{subcaption}
\usepackage[font={small,it}]{caption}
\usepackage[T1]{fontenc}
\usepackage{url}
\usepackage{enumerate}
\usepackage{hyperref} 
\graphicspath{{./figures/}} 
\usepackage{caption}
\usepackage{stfloats} 
\usepackage[blocks]{authblk}
\usepackage{epsfig}
\usepackage{latexsym}
\usepackage{soul}
\newtheorem{lemma}{Lemma}
\newtheorem{definition}{Definition}
\newtheorem{theorem}{Theorem}

\newtheorem{claim}{Claim}

\newtheorem{example}{Example}
\newtheorem{remark}{Remark}
\newtheorem{conjecture}{Conjecture}

\newcommand{\upperRomannumeral}[1]{\uppercase\expandafter{\romannumeral#1}}

\usepackage{lipsum}

\newcommand{\C}{\mathcal{C}}  
\newcommand{\That}{\widehat{T}} 
\newcommand{\U}{\mathbf{U}}

\newcommand{\E}{\mathbb{E}}
\newcommand{\un}{\mathbf{u}}
\newcommand{\pxju}{P_{\mathbf{X}_{J}|\mathbf{U}}(\mathbf{x}_{J}|\mathbf{u})}
\newcommand{\pxjum}{P_{\mathbf{X}_{J}|\mathbf{U}}(\mathbf{x}_{J}|\mathbf{u}(m))}
\newcommand{\PP}{\mathbb{P}}

\newcommand{\wjc}{\widehat{\J}^c}

\newcommand{\x}{\mathbf{x}}
\newcommand{\X}{\mathbf{X}}
\newcommand{\A}{\mathcal{A}}
\newcommand{\T}{\mathcal{T}}
\newcommand{\Y}{\mathbf{Y}}
\newcommand{\y}{\mathbf{y}}
\newcommand{\J}{J}
\newcommand{\G}{\mathcal{G}}
\newcommand{\B}{\mathcal{B}}
\newcommand{\V}{\mathbb{V}}

\newcommand{\xj}{\mathbf{x}_{J}}

\newcommand{\Xj}{\mathbf{X}_{J}}

\newcommand{\pin}{P^{\text{\text{inn}}}_{\mathbf{X}_{J}}}    
\newcommand{\ph}{\widehat{P}_{\mathbf{X}_{J}}} 
\newcommand{\pixj}{P^{\text{\text{inn}}}_{\mathbf{X}_{J}}(\mathbf{x}_{J})}   
\newcommand{\phxj}{\widehat{P}_{\mathbf{X}_{J}}(\mathbf{x}_{J})}  
\newcommand{\AXj}{\mathcal{A}^{n,\gamma}_{X_{J}}}  
\newcommand{\AU}{\mathcal{A}^{n,\gamma}_{U\mathbf{x}_{J}}}  
\newcommand{\TU}{\mathcal{T}_{\mathbf{U}\mathbf{x}_{J}}}

\begin{document}
%
\title{Stealthy Communication over Adversarially Jammed Multipath Networks}
%
%
%

\author{Jianhan Song,
	Qiaosheng Zhang,
	Swanand Kadhe, Mayank Bakshi,
	Sidharth Jaggi
	\thanks{J. Song is with the Department
		of ECE, University of Texas at Austin (jianhansong@utexas.edu). Q. Zhang is with the Department
		of ECE, National University of Singapore (elezqiao@nus.edu.sg). S. Kadhe is with the Department of EECS, University of California, Berkeley (swanand.kadhe@berkeley.edu). M. Bakshi is with the Institute of Network Coding, Chinese University of Hong Kong (mayank@inc.cuhk.edu.hk). S. Jaggi is with the Department of Information Engineering, Chinese University of Hong Kong, and School of Mathematics, University of Bristol (sid.jaggi@bristol.ac.uk). A preliminary version of this work~\cite{song2018multipath} was presented in part at the 2018 IEEE International Symposium on Information Theory (ISIT), USA.} 
}

\maketitle

\begin{abstract}
We consider the problem of \emph{stealthy communication} over a \emph{multipath network} in the presence of an active adversary. The multipath network consists of multiple parallel noiseless links, and the adversary is able to eavesdrop and jam a subset of links. We consider two types of jamming---\emph{erasure jamming} and \emph{overwrite jamming}. We require the communication to be both stealthy and reliable, i.e., the adversary should be unable to detect whether or not meaningful communication is taking place, while the legitimate receiver should reconstruct any potential messages from the transmitter with high probability simultaneously. We provide inner bounds on the stealthy capacities under both adversarial erasure and adversarial overwrite jamming.
\end{abstract}

\begin{IEEEkeywords}
Stealthy communication, Low probability of detection, Adversarial jamming, Information-theoretic security.
\end{IEEEkeywords}

%
\IEEEpeerreviewmaketitle

\section{Introduction}
Suppose an activist (Alice) {\it occasionally} wishes to communicate with a news agency, say BBC (Bob), and can use several social media accounts she has to do so. However, the government James is eavesdropping on some of these accounts (Alice and Bob do not know which ones), and is able to jam (i.e., erase or corrupt) information on these. The goal is to ensure that (i) the activist Alice can communicate with the BBC Bob even if the government James attempts to disrupt communication, and (ii) Alice's communication should be stealthy---any communication posted on the social media that James observes should be explainable as ``innocent behaviour''.

The classical information-theoretic security problem aims to hide the {\it content} of communication. However, in certain scenarios the mere {\it fact} that communication is taking place should also be hidden. Stealthy communication, first studied in~\cite{hou2014effective} for Discrete Memoryless Channels (DMCs), requires that the transmitter Alice should be able to reliably communicate with the legitimate receiver Bob, and simultaneously ensure the communication is undetectable by a malicious adversary James. The work~\cite{kadhe2014reliable} generalized the communication medium from classical DMCs to networks, and particularly studies stealthy communication over a noiseless {\it multipath network} wherein James is able to eavesdrop on a subset of links.

Stealthy communication is closely related to the well-studied {\it covert communication} problem. The major difference lies in the assumptions on the {\it innocent distribution} (when no communication happens)---covert communication requires that, under innocent transmission, the channel inputs must be the ``zero symbols'', while stealthy communication allows the inputs to follow a non-zero innocent distribution. Prior works have investigated the covert communication problem under different settings, including additive white Gaussian noise (AWGN) channels~\cite{BasGT:12a, wang2016fundamental,yan2019gaussian}, DMCs~\cite{7407378,wang2016fundamental, tahmasbi2018first}, binary symmetric channels (BSCs)~\cite{CheBJ:13}, multiple-access channels~\cite{arumugam2019covert}, broadcast channels~\cite{arumugam2019embedding, tan2018time}, compound DMCs~\cite{ahmadipour2019covert}, continuous-time channels~\cite{sobers2017covert,wang2018continuous,wang2018covert, zhang2019undetectable}, quantum channels~\cite{bash2015quantum, sheikholeslami2016covert, wang2016optimal}, {\it etc}. In particular, instead of the broadly studied {\it random noise} channels, the work~\cite{ZhangBJ:17} shifted the focus to the {\it adversarial noise} channels, i.e., the channel between Alice and Bob can be maliciously jammed by James, and the coding scheme there should be resilient to every possible (including the worst) jamming strategy induced by James.

This paper builds upon the insights obtained in ~\cite{kadhe2014reliable, ZhangBJ:17}. Suppose Alice and Bob communicate over a multipath network, which consists of $C$ parallel noiseless links. Unlike~\cite{kadhe2014reliable} wherein James is only able to eavesdrop on a subset of links {\it passively}, this work considers the situation in which James also has the ability to jam the same subset of links to disturb any potential communication (even if he cannot detect the existence of communication), based on his knowledge about the communication scheme used by Alice and Bob. When Alice does not wish to communicate with Bob, her transmissions on the $C$ links are sampled according to an innocent distribution (known {\it a priori} to Bob and James). When she is communicating with Bob, her transmissions are chosen from a public codebook. In both scenarios, James is able to control (eavesdrop on/jam) at most $Z$ out of $C$ links, but which subset of links is controlled is not known to Alice and Bob.

James first estimates whether or not Alice is transmitting by observing the transmission patterns on the links he controls. The stealth is measured via a {\it hypothesis-testing metric}---the communication is deemed to be stealthy if regardless of James' estimator, his probability of {\it false alarm} plus his probability of {\it missed detection} always approaches one asymptotically. Afterwards, on the basis of his observations and his prior knowledge about the communication scheme, James tries to adversarially jam the links he controls. We consider two types of jamming---{\it erasure jamming} and {\it overwrite jamming}. Erasure jamming means that James can only erase everything on the links he controls, while overwrite jamming allows him to replace the original transmission with his carefully designed transmission patterns. Under both erasure and overwrite jamming, we show that stealthy communication with positive rate is achievable. 

\subsection{Comparison with Related Work}

Since stealthy communication allows a non-degenerate innocent distribution, the throughputs with guarantees on both stealth and reliability, in this work and also in~\cite{hou2014effective,kadhe2014reliable}, scale linearly in the blocklength. This is in contrast to covert communication wherein one can only transmit $\mathcal{O}(\sqrt{n})$ bits covertly and reliably over $n$ channel uses. Another, somewhat technical difference, is that in our setup, the channel from Alice to James is not known {\it a priori} to Alice and Bob because of James' flexibility in choosing which subset of $Z$ links to sit on. Stealthy communication over multipath networks is also studied in~\cite{kadhe2014reliable}; however, the adversary there is passive. Furthermore, we point out that the functionalities of the adversary in this work is fundamentally different from the {\it uninformed jammer} considered in~\cite{sobers2017covert}, wherein the jammer is present to help Alice and Bob by sending ``artificial noise'' to the eavesdropper.

Another field that is closely related to stealthy communication is the {\it steganography} problem, in which Alice aims to convey a message to Bob by concealing it into the {\it covertext} (i.e., the innocent transmissions when Alice is inactive), and the adversary who has noiseless access to the {\it stegotext} (i.e., the transmissions from Alice to Bob) should not be able to detect the existence of the hidden message. Unlike this work, the adversary in the steganography problem usually has {\it noiseless} observations of the transmissions, and most works assume that {\it shared key} between Alice and Bob is available. An information-theoretic model of steganography is first proposed by Cachin~\cite{cachin1998information}, and several follow-up works~\cite{moulin2003information,moulin2004new,wang2007capacity,wang2008perfectly,mittelholzer1999information} also take the active jamming into account. However, most of these works (except~\cite{moulin2004new, mittelholzer1999information}) focus on the {\it memoryless attack} or {\it blockwise memoryless attack} (i.e., the attack channel designed by the adversary is essentially a memoryless channel), and they usually impose \emph{distortion constraints} on the attack channel. More importantly, the schemes in all these works rely critically on the shared key between Alice and Bob. On the contrary, this work does not require the distortion constraints, the shared key, and the channel to be memoryless---our scheme works as long as the adversary's channel is worse than Bob's channel, and the analysis relies on the imperfection of the adversary's observations.      

Reliable communication (without stealth constraints) over a multipath network in the presence of an adversary has been well-studied in the past~\cite{jaggi2005correction, zhang2015talking, zhang2015coding, kadhe2015reliable}. The work~\cite{jaggi2005correction} first shows that as long as $Z < C/2$, Alice and Bob can fully utilize the rest of links to communicate, under both  erasure and overwrite jamming. Robustness against erasure jamming is relatively straightforward while robustness against overwrite jamming requires non-trivial coding schemes (such as {\it pairwise hashing}~\cite{jaggi2005correction}). Similar results are obtained in this work while also taking stealth into account.


\subsection{Our Contributions and High-level Intuition}

Our schemes are the first that can attain two simultaneous goals---ensure stealth (i.e., James cannot infer whether or not meaningful communication is occurring) and in parallel also ensure robustness to jamming (i.e., James is unable to corrupt meaningful communication if it is happening). Note that James is quite strong---he is computationally unbounded, knows \emph{a priori} the communication scheme (including the encoder, decoder, and codebook) that would be used if meaningful communication were indeed happening, and is able to eavesdrop on any subset of links of size at most $Z$ and base his jamming strategy on what he sees (even if he is unable to detect communication happening). However, Alice and Bob do not know the subset James controls as well as  the jamming strategy he uses.

Under erasure jamming, the channel between Alice and James can be viewed as an aggregation of all the links controlled by James, while the channel between Alice and Bob can be viewed as an aggregation of the complement of these links (since James erases everything on the links he controls). The stealth constraint imposes a lower bound on the rate (as a consequence of the \emph{channel resolvability}~\cite{han1993approximation,7407378}), while the reliability constraint imposes an upper bound (as a consequence of the channel coding theorem). As is standard in wiretap secrecy problems, we create an artificial noisy channel at the encoder (which may hurt James more than Bob) in our scheme to obtain a higher rate compared to a relatively straightforward approach.

Coding against an overwrite adversary is significantly more non-trivial since James can use any jamming strategy which is unknown to Alice and Bob. In this work we develop a coding scheme with positive rate that is resilient to every (including even the worst-case) possible jamming strategy. The essences of our proof lie in Lemma~\ref{lemma:ratio} presented in Section~\ref{sec:overwrite} and a proper use of the \emph{McDiarmid's inequality}~\cite{mcdiarmid1989method}.


While the focus of this work is on robustness to active jamming, it has not escaped our attention that composing our schemes with well-known techniques in
the information-theoretic literature allows us to get schemes that are secure against both information leakage and active jamming attacks in this stealthy communication setting. A full characterization of this communication setting with trifold objectives is a source of ongoing investigation.

\section{Model} \label{sec:model}
Random variables and their realizations are respectively denoted by uppercase letters and lowercase letters, e.g., $X$ and $x$. Sets
are denoted by calligraphic letters, e.g., $\mathcal{X}$. Vectors of length-$n$ are denoted by boldface letters, e.g., $\X$ and $\x$. If the single-letter distribution on $X$ is $P_X$, then the corresponding $n$-letter product distribution $\prod_{i=1}^n P_X$ is denoted by $P_{\X}$.  Throughout this paper we use {\it asymptotic notations}~\cite[Ch. 3.1]{leiserson2001introduction} to describe the limiting behaviour of functions. 

\begin{figure}
	\begin{center}
		\includegraphics[scale=0.58]{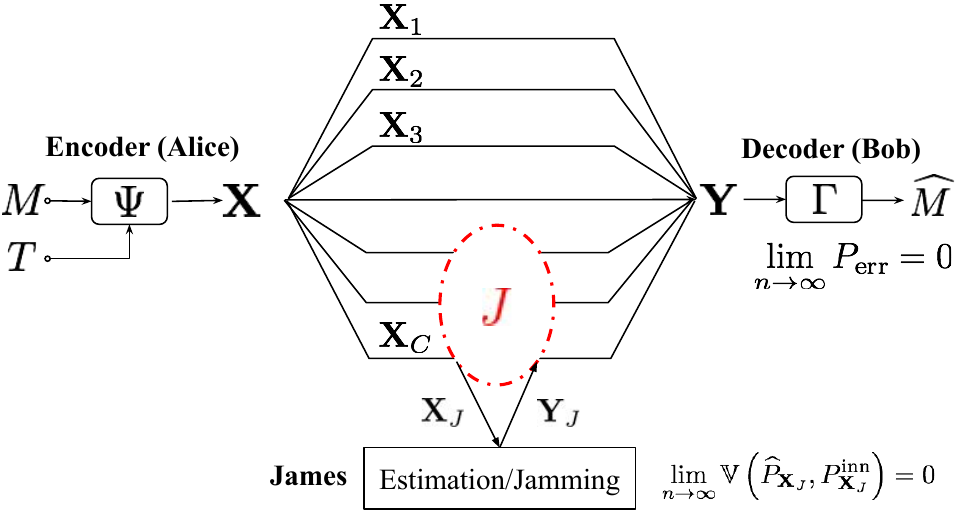}
		\caption{System diagram} \label{fig:system}
	\end{center}
\vspace{-30pt}
\end{figure}

The multipath network consists of $C$ parallel links $L_1, L_2, \ldots, L_C$, each link $L_i$ carries a symbol from the alphabet $\mathcal{X}_i$ per time instant. The alphabet for all the links taken together is denoted by $\mathcal{X} \triangleq \prod_{i=1}^C \mathcal{X}_i.$ 
Alice's transmission status is denoted by $T \in \{0,1\}$: $T=0$ if Alice is innocent, whereas $T = 1$ if Alice is active. The message $M$ is either $0$ (if Alice is innocent) or uniformly distributed over $\{1,2,\ldots, N\}$ (if Alice is active). Note that no prior distribution is assigned to $T$ and only Alice knows the values of $T$ and $M$ {\it a priori}. Let $n$ be the blocklength (number of time instants). The length-$n$ vector transmitted on the $j$-th link is denoted by $\x_j$, and the collection of vectors on $C$ links is denoted by $\x = [\x_1^T \ \x_2^T \ldots \x_C^T]^T$. Note that $\x$ can also be viewed as a length-$n$ vector over $\mathcal{X}$. The system diagram is illustrated in Figure~\ref{fig:system}.

\subsubsection{Innocent distribution} When Alice is innocent ($T=0$), at each time instant $t$ ($1 \le t \le n$), an innocent transmission pattern on the $C$ links is sampled according to the time-independent innocent distribution $P^{\text{inn}}_X \in \mathcal{P}(\mathcal{X})$, where $\mathcal{P}(\mathcal{X})$ denotes the set of all distributions on $\mathcal{X}$. For any subset $\J \subseteq \{L_1, L_2, \ldots, L_C\}$, the {\it marginal innocent distribution} is denoted by $P^{\text{inn}}_{X_{\J}}$. Over $n$ time instants, the corresponding {\it n-letter innocent distribution} and {\it n-letter marginal innocent distribution} (for subset $J$) are product distributions with the form 
\begin{align*}
&P^{\text{inn}}_{\X} \triangleq \prod_{t=1}^n P^{\text{inn}}_X, \quad P^{\text{inn}}_{\X_\J} \triangleq \prod_{t=1}^n P^{\text{inn}}_{X_{\J}}.
\end{align*}

\subsubsection{Encoder} Alice's encoder $\Psi(.,.)$ takes the transmission status $T$ and the message $M$ as input, and outputs a length-$n$ vector $\X$. If $T = 0$ (thus $M = 0$), the encoder $\Psi(0,0)$ outputs a vector $\X$ according to the innocent distribution $P^{\text{inn}}_{\X}$. If $T = 1$ and message $m$ is transmitted, the encoder $\Psi(1,m)$ outputs the corresponding length-$n$ vector $\x(m)$ for transmission. The rate is defined as 
\begin{align}
R \triangleq \frac{\log N}{n}. 
\end{align}
Under overwrite jamming, the codebook is the collection of $N$ length-$n$ vectors $\{\x(m)\}_{m=1}^N$ over $\mathcal{X}^n$; while under erasure jamming, the codebook is the collection of $N$ length-$n$ vectors $\{\un(m)\}_{m=1}^N$ over $\mathcal{U}^n$ (as detailed in Section~\ref{sec:erasure}, we first map the message $m$ to the codeword $\un(m)$ and then stochastically map $\un(m)$ to $\x(m)$ for transmission).
We assume that the codebook is known to all parties including the adversary.

\subsubsection{Active distribution} The active distribution, which is averaged over the codebook, is denoted by $\widehat{P}_{\X}$. Similarly, for any subset $\J \subseteq \{L_1, L_2, \ldots, L_C\}$, the {\it marginal active distribution} is denoted by $\widehat{P}_{\X_\J}$.

\subsubsection{James' estimation and jamming} The adversary James knows \emph{a priori} the communication scheme (including the encoder, decoder, and codebook) that would be used if meaningful communication were indeed happening. Let $\mathfrak{J}$ be the class of all possible subsets of size at most $Z$, i.e.,
$\mathfrak{J} \triangleq \{J \subseteq \{L_1, L_2, \ldots, L_C\}: |J| \le Z\}.$
James is able to control \emph{any} subset $\J \in \mathfrak{J}$, and his choice is unknown to both Alice and Bob. Moreover, James is also assumed to be computationally unbounded.
 On the basis of his observations on the subset he controls,  James estimates Alice's transmission status $T$, and also non-causally jams the subset to prevent reliable communication irrespective of his estimation.

\noindent{\it \underline{Estimation:}} James' estimator $\Phi(.)$ outputs a single bit $\widehat{T} = \Phi(\X_\J)$ to estimate Alice's transmission status $T$. We respectively defined the probability of false alarm and the probability of missed detection of an estimator $\Phi$ as 
$\alpha(\Phi) \triangleq \PP(\That = 1| T = 0)$ and $\beta(\Phi) \triangleq \PP(\That = 0| T = 1)$.
We use a  hypothesis testing metric to measure the stealth.
\begin{definition}[Stealthy Communication]
	The communication is said to be stealthy if 
	\begin{align}
	\lim_{n \to \infty} \min_{\Phi}\{\alpha(\Phi) + \beta(\Phi) \} = 1.
	\end{align}
\end{definition}
In other words, stealthy communication requires that regardless of which estimator $\Phi$ is chosen, $\alpha(\Phi)+\beta(\Phi)$ should always approach one as $n$ tends to infinity. Note that a na\"ive estimator $\tilde{\Phi}$ (which always outputs $\widehat{T} = 0$ or $\widehat{T}=1$) also guarantees $\alpha(\tilde{\Phi}) + \beta(\tilde{\Phi}) = 1$. Therefore, the definition for stealthy communication implies that James' optimal estimator (denoted by $\Phi^\ast$) cannot be much better than the na\"ive estimator $\tilde{\Phi}$. A classical result on hypothesis testing~\cite{lehmann2006testing} shows that the optimal estimator $\Phi^{\ast}$ satisfies 
$\alpha(\Phi^{\ast})+\beta(\Phi^{\ast}) = 1 - \V(\widehat{P}_{\X_{\J}}, P^{\text{inn}}_{\X_{\J}}),$ 
where $\V(\widehat{P}_{\X_{\J}}, P^{\text{inn}}_{\X_{\J}}) \triangleq \frac{1}{2}\sum_{\x_\J}|\widehat{P}_{\X_{\J}}(\x_\J) - P^{\text{inn}}_{\X_{\J}}(\x_\J)|$ is the {\it variational distance} between the marginal active distribution and the marginal innocent distribution. To prove the communication is stealthy, it is equivalent to showing that for every $J \in \mathfrak{J}$,
\begin{align}
\lim_{n \to \infty}\V(\widehat{P}_{\X_{\J}}, P^{\text{inn}}_{\X_{\J}}) = 0.
\end{align}

\noindent{\it \underline{Jamming:}} James is also able to maliciously jam the set $J$ he controls. Under erasure jamming, the transmission $\x_\J$ is completely replaced by the erasure symbols `$\perp$', while under overwrite jamming, $\x_\J$ is replaced by a carefully designed $\y_\J$. In particular, James is able to choose the jamming vector $\y_\J$ stochastically according to {\it any} conditional distribution $W_{\Y_\J|\X_\J,\C}$, since he knows $\x_{\J}$ and the codebook. Note that Alice and Bob do not know James' jamming strategy.

\subsubsection{Decoder} Bob receives $\y$ through the multipath network. 
\begin{enumerate}
	\item Under erasure jamming, $\y_{\J^c}=\x_{\J^c}$ on the subset $\J^c$ (where $\J^c$ denotes the complement of set $\J$), and $\y_{\J}$ equals the erasure symbols `$\perp$' on the subset $\J$.
	\item Under overwrite jamming, $\y_{\J^c}=\x_{\J^c}$ on $\J^c$, while $\y_\J$ is arbitrarily chosen by James.
\end{enumerate}
Note that Bob can easily figure out the subset $\J$ under erasure jamming due to the appearance of `$\perp$', while it is not the case under overwrite jamming.
Bob reconstructs the message $\widehat{M}$ by applying his {\it decoding function} $\Gamma(.)$ to his observation. The probabilities of error under erasure and overwrite jamming are respectively defined as
\begin{align*}
&P_{\text{err}}^{\perp}(\Psi, \Gamma) \triangleq \max_{\J \in \mathfrak{J}} \sum_{t \in \{0,1\}} \PP(\widehat{M} \ne M | T=t), \ \ P_{\text{err}}^{\text{ow}}(\Psi, \Gamma) \triangleq \max_{\J \in \mathfrak{J}} \max_{W_{\Y_\J|\X_\J,\C}} \sum_{t \in \{0,1\}} \PP(\widehat{M} \ne M | T=t).
\end{align*}

\subsubsection{Achievable rate} A rate $R$ is said to be {\it achievable under erasure jamming (resp. achievable under overwrite jamming)} if there exists an infinite sequence of codes $(\Psi_n, \Gamma_n)$ such that each code in the sequence has rate at least $R$, and ensures $\lim_{n \to \infty}\V(\widehat{P}_{\X_{\J}}, P^{\text{inn}}_{\X_{\J}}) = 0$ for every $J \in \mathfrak{J}$ and $\lim_{n \to \infty}P_{\text{err}}^{\perp}(\Psi_n, \Gamma_n) = 0$ ({\it resp.} $\lim_{n \to \infty}P_{\text{err}}^{\text{ow}}(\Psi_n, \Gamma_n) = 0$).

\section{Main results} \label{sec:result}

To facilitate the statement of our results, we first define an optimization problem (A), which includes an auxiliary random variable $U$, for a fixed innocent distribution $P^{\text{inn}}_X$ and a non-negative integer $Z < C/2$ as follows:   
\begin{align}
(A) \ \ & \sup_{P_{U},P_{X|U}}
& &\min_{\J \in \mathfrak{J}} I(U;X_{\J^c}) \notag  \\
& \text{subject to} 
& & P^{\text{inn}}_{X_{\J}} = \sum_{u} P_{U} \cdot P_{X_{\J}|U}, \ \forall \J \in \mathfrak{J}, \label{eq:con3}\\
&&& \max_{\J \in \mathfrak{J}} I(U;X_{\J}) < \min_{\J \in \mathfrak{J}} I(U;X_{\J^c}). \label{eq:con4}
\end{align}
The optimal value of (A) is denoted by $\bar{K}(P^{\text{inn}}_X,Z)$. Consider another optimization
\begin{align}
(B)\  \ & \sup_{P_{X}}
& &\min_{\J \in \mathfrak{J}} H(P_{X_{\J^c}}) \notag \\
& \text{subject to} 
& & P^{\text{inn}}_{X_{\J}} = P_{X_\J}, \ \forall \J \in \mathfrak{J}, \label{eq:con1} \\
&&& \max_{\J \in \mathfrak{J}} H(P_{X_{\J}}) < \min_{\J \in \mathfrak{J}} H(P_{X_{\J^c}}), \label{eq:con2}  
\end{align}
and let the optimal value be $\underline{K}(P^{\text{inn}}_X,Z)$. It is worth noting  that $\underline{K}(P^{\text{inn}}_X,Z)$ is always bounded from above by $\bar{K}(P^{\text{inn}}_X,Z)$, since (A) is equivalent to (B) by restricting $U=X$. In the following, we provide an example showing that $\underline{K}(P^{\text{inn}}_X,Z)$ is sometimes strictly smaller than $\bar{K}(P^{\text{inn}}_X,Z)$. 

\begin{example} \label{example:1}
Suppose the multipath network contains three links ($C = 3$), James is able to arbitrarily control one link ($Z = 1$), and the alphabet of each link is binary, i.e., $\mathcal{X}_1 = \mathcal{X}_2 = \mathcal{X}_3 = \{0,1\}$. Let the innocent distribution $P^{\text{inn}}_{X}$ be a product distribution, i.e., $P^{\text{inn}}_X = P^{\text{inn}}_{X_1}P^{\text{inn}}_{X_2}P^{\text{inn}}_{X_3}$, with $P^{\text{inn}}_{X_1}(0) = P^{\text{inn}}_{X_2}(0) = 0.1$ and $P^{\text{inn}}_{X_3}(0) = 0.5$. We first show that the optimization (B) is infeasible. This is because for all $P_X$ satisfying the first constraint of optimization (B), the second constraint of optimization (B) cannot be satisfied since $H(P_{X_1X_2}) \le H(P_{X_1}) + H(P_{X_2}) = H(0.1)+H(0.1) = 0.938 < 1 = H(P_{X_3})$. Therefore, $\underline{K}(P^{\text{inn}}_{X},Z) = 0$ in this setting.  

By introducing an auxiliary random variable $U$ with $\mathcal{U} = \{0,1\}$, the optimal value $\bar{K}(P^{\text{inn}}_{X},Z)$ of optimization (A) becomes non-zero. We choose $P_U(0)=0.2$ and the conditional probability $P_{X|U}=P_{X_1|U}P_{X_2|U}P_{X_3|U}$, where $P_{X_1|U}, P_{X_2|U}, P_{X_3|U}$ are given in the following table.
\begin{center}
	\scriptsize
	\begin{tabular}{|c|c|c|c|c|c|c|}
		\hline
		& \multicolumn{2}{c|}{$P_{X_1|U}$} & \multicolumn{2}{c|}{$P_{X_2|U}$} & \multicolumn{2}{c|}{$P_{X_3|U}$} \\ \hline
		& $X_1=0$        & $X_1=1$       & $X_2=0$        & $X_2=1$       & $X_3=0$        & $X_3=1$       \\ \hline
		$U=0$ &    $0.5$ &  $0.5$   &   $0.5$   &   $0.5$ &   $0.9$   &  $0.1$   \\ \hline
		$U=1$ &    $0$  &  $1$    &   $0$    &   $1$   &   $0.4$     &   $0.6$   \\ \hline
	\end{tabular} 
\end{center}
One can verify that such choices of $P_U$ and $P_{X|U}$ satisfy the first constraint of optimization (A). Moreover, we have $\max_{J \in \mathfrak{J}} I(U;X_{J}) = I(U;X_1) = 0.269$, and 
\begin{align}
\min_{J \in \mathfrak{J}} I(U;X_{J^c}) &= I(U;X_2X_3) \stackrel{\mathrm{(a)}}{=} I(U;X_2)+I(U;X_3) - I(X_2;X_3) = 0.34, \notag 
\end{align}
where (a) follows since $X_2 - U - X_3$ forms a Markov chain. Thus, the second constraint of optimization (A) is also satisfied. Therefore, we know that $\bar{K}(P^{\text{inn}}_{X},Z)$ is at least $0.34$.  
\end{example}

As is usual in wiretap secrecy problems, Theorem~\ref{thm:erasure} below shows that a higher rate $\bar{K}(P^{\text{inn}}_X,Z)-\epsilon$ is achieved by introducing an auxiliary variable $U$.

\begin{theorem}[Erasure jamming] \label{thm:erasure}
	For any $P^{\text{inn}}_X$ and non-negative integer $Z < C/2$, the rate 
	$R = \bar{K}(P^{\emph{inn}}_X,Z) - \epsilon$ is achievable under erasure jamming for sufficiently small $\epsilon >0$.
\end{theorem}

Lemma~\ref{lemma:cardinality} below provides a bound on the cardinality of the random variable $U$. The proof relies on standard cardinality bound arguments~\cite{el2011network} and is deferred to  Appendix~\ref{appendix:cardinality}
\begin{lemma}[Cardinality Bound] \label{lemma:cardinality}
	Given any feasible random variable $(U, X)$ in optimization (A), there exists a feasible $(U',X)$ with $|\mathcal{U}'| \leq |\mathcal{X}| + 2|\mathfrak{J}| -1$ that yields the same objective value.
\end{lemma}

Compared to erasure jamming, dealing with overwrite jamming is much more challenging due to the fact that James, knowing Alice's codebook, may attempt to ``spoof'' Alice's transmissions. Bob's decoder should be robust to any jamming strategy (or any conditional distribution) $W_{\Y_\J|\X_{\J},\C}$, including the one that maximizes his probability of decoding error. However, our next result shows that stealthy communication with positive rate is still possible. 

\begin{theorem}[Overwrite jamming] \label{thm:overwrite}
	For any $P^{\text{inn}}_X$ and non-negative integer $Z < C/2$, the rate $R = \underline{K}(P^{\emph{inn}}_X,Z)-\epsilon$ is achievable under overwrite jamming for sufficiently small $\epsilon >0$.
\end{theorem}

In addition to the achievability results presented in Theorems~\ref{thm:erasure} and~\ref{thm:overwrite}, Theorem~\ref{thm:converse} below also provides upper bounds for both erasure jamming and overwrite jamming when $Z < C/2$. 

\begin{theorem}[Upper bounds for $Z < C/2$]\label{thm:converse} Under erasure jamming (resp. overwrite jamming), we consider a sequence of codes with increasing blocklength $n$ such that $\varepsilon_n \triangleq P_{\emph{err}}^{\perp}$ (resp. $\varepsilon_n \triangleq P_{\emph{err}}^{\emph{ow}}$) and $\delta_n \triangleq \max_{J \in \mathfrak{J}} \V(\widehat{P}_{\X_{\J}}, P^{\text{inn}}_{\X_{\J}})$. If $\lim_{n \to \infty} \varepsilon_n = \lim_{n \to \infty} \delta_n = 0$, we have that for any $\epsilon \in (0,1)$,
\begin{align*}
\lim_{n \to \infty} R = \lim_{n \to \infty} \frac{\log N}{n} \le \sup_{P_{X}: \V(P_{X_J}, P^{\text{inn}}_{X_J}) \le \epsilon, \forall J \in \mathfrak{J} } \min_{J \in \mathfrak{J}} H(P_{X_{J^c}}).
\end{align*}	
\end{theorem}

\begin{proof}
	For any sequence of codes satisfying $\lim_{n \to \infty} \varepsilon_n = \lim_{n \to \infty} \delta_n = 0$, we have 
	\begin{align*}
		\log N = H(M) &= I(M;\widehat{M}) + H(M|\widehat{M}) \\
		&\stackrel{\text{(a)}}{\le} I(M;\widehat{M}) + n \varepsilon'_n \\
		&\stackrel{\text{(b)}}{\le} \min_{J \in \mathfrak{J}} I(\X; \X_{J^c}) + n \varepsilon'_n \\
		&\le \min_{J \in \mathfrak{J}} \sum_{i=1}^n I(X_i; (X_i)_{J^c}) + 	n \varepsilon'_n,
	\end{align*} 
where (a) follows from Fano's inequality and $\varepsilon'_n$ is a sequence that  depends on $\varepsilon_n$ and satisfies $\lim_{n \to \infty} \varepsilon'_n = 0$, (b) follows from data processing inequality, the fact that Bob can observe $\Y_{\J^c} = \X_{J^c}$ noiselessly, and the fact that James can choose any $J \in \mathfrak{J}$ to minimize the mutual information $I(\X; \X_{J^c})$. Recall that for any $J \in \mathfrak{J}$, the $n$-letter distribution $\widehat{P}_{\X_{\J}}$ (induced by the code) satisfies $\V(\widehat{P}_{\X_{\J}}, P^{\text{inn}}_{\X_{\J}}) \le \delta_n$, and let $(\widehat{P}_{\X_{\J}})_i$, for $i \in \{1,2,\ldots, n\}$, be the marginal distributions of $\widehat{P}_{\X_{\J}}$. We then have that for each $J \in \mathfrak{J}$,
\begin{align*}
\V((\widehat{P}_{\X_{\J}})_i, P^{\text{inn}}_{X_{\J}}) &= \frac{1}{2}\sum_{x_J^{(i)}}\left|(\widehat{P}_{\X_{\J}})_i(x_J^{(i)}) - P^{\text{inn}}_{X_{\J}}(x_J^{(i)}) \right| \\
&\stackrel{\text{(c)}}{=} \frac{1}{2}\sum_{x_J^{(i)}}\left|\sum_{\x_J^{(-i)}} (\widehat{P}_{\X_{\J}})_i(x_J^{(i)}) \frac{\widehat{P}_{\X_{\J}}(x_J^{(i)},\x_J^{(-i)})}{(\widehat{P}_{\X_{\J}})_i(x_J^{(i)})} - \sum_{\x_J^{(-i)}} P^{\text{inn}}_{\X_{\J}}(x_J^{(i)},\x_J^{(-i)}) \right| \\
&\le \frac{1}{2}\sum_{x_J^{(i)}} \sum_{\x_J^{(-i)}} \left| (\widehat{P}_{\X_{\J}})_i(x_J^{(i)}) \frac{\widehat{P}_{\X_{\J}}(x_J^{(i)},\x_J^{(-i)})}{(\widehat{P}_{\X_{\J}})_i(x_J^{(i)})} - P^{\text{inn}}_{\X_{\J}}(x_J^{(i)},\x_J^{(-i)}) \right| \\
&= \V(\widehat{P}_{\X_{\J}}, P^{\text{inn}}_{\X_{\J}}) \le \delta_n,
\end{align*}  
where $\x_J^{(-i)} \triangleq [x_J^{(1)},\ldots, x_J^{(i-1)}, x_J^{(i+1)}, \ldots, x_J^{(n)}]$ and (c) holds since $P^{\text{inn}}_{\X_\J} = \prod_{t=1}^n P^{\text{inn}}_{X_{\J}}$.
That is, any sequence of codes satisfying $\max_{J \in \mathfrak{J}} \V(\widehat{P}_{\X_{\J}}, P^{\text{inn}}_{\X_{\J}}) = \delta_n$ has  the property that for any $J \in \mathfrak{J}$, the marginal distribution $(\widehat{P}_{\X_{\J}})_i$ is close the the innocent distribution $P^{\text{inn}}_{X_{\J}}$, i.e., $\V((\widehat{P}_{\X_{\J}})_i, P^{\text{inn}}_{X_{\J}}) \le \delta_n$. Thus, the mutual information $I(X_i; (X_i)_{J^c})$ for $i \in \{1,2,\ldots, n\}$ satisfies
\begin{align*}
&I(X_i; (X_i)_{J^c}) \le \sup_{P_{X}: \V(P_{X_J}, P^{\text{inn}}_{X_J}) \le \delta_n, \forall J \in \mathfrak{J} } H(P_{X_{J^c}}) \stackrel{\text{(d)}}{\le} \sup_{P_{X}: \V(P_{X_J}, P^{\text{inn}}_{X_J}) \le \epsilon, \forall J \in \mathfrak{J} } H(P_{X_{J^c}}),
\end{align*}
where (d) holds for sufficiently large $n$ since $\epsilon \in (0,1)$ is independent of $n$. Therefore, we have
\begin{align*}
	\lim_{n \to \infty} R = \lim_{n \to \infty}\frac{\log N}{n} \le \min_{J \in \mathfrak{J}} \sup_{P_{X}: \V(P_{X_J}, P^{\text{inn}}_{X_J}) \le \epsilon, \forall J \in \mathfrak{J} } H(P_{X_{J^c}}) = \sup_{P_{X}: \V(P_{X_J}, P^{\text{inn}}_{X_J}) \le \epsilon, \forall J \in \mathfrak{J} } \min_{J \in \mathfrak{J}} H(P_{X_{J^c}}).
\end{align*}

\end{proof}

\begin{remark}
When $Z \ge C/2$, it is impossible to communicate reliably and stealthily simultaneously under both erasure jamming and overwrite jamming. To explain the rational behind the above argument, in the following we consider a concrete setting in which $C$ is even and $Z = C/2$. The analysis for the setting in which $Z > C/2$ is similar.  

\underline{(a) Erasure Jamming}: Note that James is able to choose \emph{any} subset among all subsets of size $C/2$, and Bob is required to reliably decode regardless of which subset is chosen by James. If James chooses $J = \{1,\ldots, C/2\}$, Bob is required to decode the message reliably based on his observations $[\x_{C/2+1}, \ldots, \x_{C}]$ on the subset $J^c$. Thus, if James chooses a different subset $J$ which equals $\{C/2+1,\ldots, C\}$, he observes $[\x_{C/2+1}, \ldots, \x_{C}]$ and then will also be able to reliably decode the message (since the communication scheme including the codebook is public). This implies that if the communication is reliable, it cannot be stealthy simultaneously.   

\underline{(b) Overwrite Jamming}: When James controls at least half of the links, he is at least as powerful as Alice (since Alice and Bob do not have any shared key). Whatever Alice does, James can do as well---James can pretend to be the transmitter and send a fake message to Bob (using Alice's encoder and the public codebook) on the links he controls; thus, Bob is unable to distinguish Alice's true message and James' fake message. This implies that Bob's probability of decoding error cannot be vanishing.
\end{remark}

\begin{remark}
	Under a slightly different setting in which the adversary James does not know the codebook (the other assumptions are the same), the maximum rates we achieve are still $\bar{K}(P^{\emph{inn}}_X,Z)-\epsilon$ (under erasure jamming) and $\underline{K}(P^{\emph{inn}}_X,Z)-\epsilon$ (under overwrite jamming). In fact, one of our main contribution is that the communication schemes we developed are robust to a stronger adversary (i.e., knowing the codebook) and simultaneously achieve the same rates compared to the setting with a weaker adversary (i.e., not knowing the codebook). 
\end{remark}

\section{Erasure Jamming}\label{sec:erasure}

\subsection{Achievability Scheme}
Our achievability scheme relies on a {\it random coding argument}. Let the optimal distributions in optimization (A) be $P_U$ and $P_{X|U}$.

\noindent{\it \underline{Encoder:}} We set
$R =\bar{K}(P^{\text{inn}}_X,Z) - \epsilon = \min_{\J \in \mathfrak{J}} I(U;X_{\J^c}) - \epsilon$ for some sufficiently small $\epsilon > 0$ such that $R > \max_{\J \in \mathfrak{J}} I(U;X_{\J})$, where the random variable pair $(U,X)$ is distributed according to $P_U \cdot P_{X|U}$. For each message $m \in \{1,2,\ldots,N\}$, where $N = 2^{nR}$, the codeword $\un(m)$ is generated according to the $n$-letter product distribution $P_\U \triangleq \prod_{i=1}^n P_U$. The codebook $\C$ is the collection of all codewords $\{\un(m)\}_{m=1}^N$. To transmit $m$, Alice chooses $\un(m)$ and stochastically maps $\un(m)$ to $\x(m)$ according to the $n$-letter product distribution $P_{\X|\U}(\x(m)|\un(m))$, and $\x(m)$ is then transmitted over the multipath network. 

\noindent{\it \underline{Decoder:}} Bob first determines the subset $\J$ (controlled by James) based on the erasure symbol `$\perp$', and then applies {\it typicality decoding} based on $\y_{\J^c}$. Note that $\y_{\J^c}=\x_{\J^c}$ since the subset $\J^c$ is not controlled by James. He decodes to $\widehat{T}=1$ and $\widehat{M}=m$ if there exists a unique $m$ such that $(\un(m),\y_{\J^c})$ are jointly typical, whereas $\widehat{T}=0$ and $\widehat{M}=0$ if there does not exist any $m$ such that $(\un(m),\y_{\J^c})$ are jointly typical.

\subsection{Proof Sketch of Stealth} \label{sec:sketch_stealth}
We provide a proof sketch of stealth in this subsection, and defer the detailed proof to Appendix~\ref{proof:stealth}).
To satisfy the stealth constraint, one should guarantee that no matter which subset $\J$ is controlled by James, the marginal active distribution $\widehat{P}_{\X_{\J}}$ is indistinguishable from the marginal innocent distribution $P^{\text{inn}}_{\X_{\J}}$. Note that
\begin{align}
\widehat{P}_{\X_{\J}}(\x_\J) = \sum_{m=1}^N \frac{1}{N} P_{\X_{\J}|\U}(\x_\J|\un(m)), \label{eq:hat1} \\
P^{\text{inn}}_{\X_{\J}}(\x_\J) = \sum_{\un}P_{\U}(\un)P_{\X_{\J}|\U}(\x_\J|\un). \label{eq:innocent}
\end{align}
Equation~\eqref{eq:innocent} follows from the constraint in~\eqref{eq:con3}, which ensures that the stochastic process $\sum_{u}P_U\cdot P_{X_\J|U}$ simulated by the encoder $\Psi$ is identical to the marginal innocent distribution $P^{\text{inn}}_{\X_{\J}}$.
The constraint in~\eqref{eq:con4} ensures the size of the codebook to be large enough so that with high probability the active distribution $\widehat{P}_{\X_{\J}}$ is sufficiently close to $\sum_{u}P_U\cdot P_{X_\J|U}$ --- it turns out that $R > I(U;X_{\J})$ is sufficient, as noticed in~\cite{7407378}, from a channel resolvability perspective. 
To prove it, we first denote the {\it typical set} of $X_\J$ by $\AXj$ (where $\gamma \to 0$ as $n \to \infty$), and the {\it jointly typical set} ({\it resp. joint type class}) of $U$ with respect to a typical $\x_\J$ by $\A_{U\x_\J}$ ({\it resp.} $\T_{U\x_\J}$). In the following, we drop the subscripts of $P_{\U}$ and $P_{\X_{\J}|\U}$ for notational convenience. Recall that proving stealth is equivalent to bounding the variational distance  $\V(P^{\text{inn}}_{\X_{\J}},\widehat{P}_{\X_{\J}}) = \frac{1}{2}\sum_{\xj}|\pixj - \phxj|.$
For any typical $\x_\J$, we have
\begin{align}
\left|\pixj - \phxj\right| &\stackrel{\text{(a)}}{\approx} \bigg| \sum_{\un \in \A_{U\x_\J}}P(\un)P(\xj|\un) -\sum_{m: \un(m) \in \A_{U\x_\J}}\frac{1}{N}P(\xj|\un(m)) \bigg| \notag\\
&\stackrel{\text{(b)}}{\le} \sum_{\T_{U\x_\J}} \bigg| \sum_{\un \in \T_{U\x_\J}}P(\un)P(\xj|\un) -\sum_{m: \un(m) \in \T_{U\x_\J}}\frac{P(\xj|\un(m))}{N} \bigg|\notag \\
&\stackrel{\text{(c)}}{=}\sum_{\T_{U\x_\J}}P(\xj|\un)  \bigg|P(\U \in \T_{U\x_\J}) - \frac{\left|m:\un(m) \in \T_{U\x_\J}\right|}{N} \bigg| \label{eq:14},
\end{align}
where the approximation (a) is obtained by discarding negligible atypical events (see~\eqref{eq:haha} in Appendix~\ref{proof:stealth} for a detailed calculation), (b) is obtained by dividing the typical set $\A_{U\x_\J}$ into typical type classes $\T_{U\x_\J}$, and (c) follows since $P(\x_\J|\un)$ is identical for all $\un \in \T_{U\x_\J}$. Note that 
\begin{align}
\mu \triangleq \E_{\C}\left(\left|m:\un(m) \in \T_{U\x_\J}\right|\right) &= N \cdot P(\U \in \T_{U\x_\J}),
\end{align}
which is exponentially large since $P(\U \in \T_{U\x_\J})\stackrel{\cdot}{=}2^{-nI(U;X_{\J})}$ and $\log N = nR > nI(U;X_{\J})$ (due to the code design). One can apply the {\it Chernoff bound}~\cite{chernoff1952measure} (which is provided in Appendix~\ref{appendix:chernoff}) to show that with probability at least $1-2e^{-\frac{1}{3}\mu \varepsilon_n^2}$ (i.e., super-exponentially close to one) over the code design,
\begin{align}
\Big|P(\U \in \T_{U\x_\J})- \frac{|m: \un(m)\in \T_{U\x_\J}|}{N}\Big| \le \varepsilon_n P(\U \in \T_{U\x_\J}), \label{eq:17}
\end{align}
where $\varepsilon_n \to 0$ as $n \to \infty$. Finally, by substituting~\eqref{eq:17} for~\eqref{eq:14}, and by taking a union bound over exponentially many $\T_{U\x_\J}$ and $\x_\J$, we prove that $\V(\pin, \ph ) \le f(\epsilon_n)$ with high probability for some function $f(\cdot)$, where $f(\varepsilon_n) \to 0$ as $n \to \infty$. Finally, note that the above analysis holds for every possible subset $\J \in \mathfrak{J}$ that James may choose, since the rate $R > \max_{\J \in \mathfrak{J}} I(U;X_{\J})$.

\subsection{Proof of Reliability}

To guarantee reliability, we note that the effective channel between Alice and Bob is $P_{X_{\J^c}|U}$ under erasure jamming, since Bob observes $\Y_{\J^c} = \X_{\J^c}$ noiselessly. Recall that our achievability relies on a random coding argument with input distribution $P_U$ and an effective channel $P_{X_{\J^c}|U}$. Since the rate $R$ is smaller than $\min_{\J \in \mathfrak{J}}I(U;X_{\J^c})$, the random coding argument directly implies that with high probability over the code design, the probability of error tends to zero as $n$ tends to infinity, regardless of which subset $\J \in \mathfrak{J}$ is chosen by James.

\section{Overwrite Jamming}\label{sec:overwrite}
We first highlight two challenges for reliable decoding under overwrite jamming: 
{(i)} In contrast to erasure jamming, it is not trivial for Bob to figure out which subset $\J \in \mathfrak{J}$ is controlled by James. In fact, our coding scheme described below requires Bob to try every possible choice of $\J$.
{(ii)} Though James can only control set $\J$, he is not ``completely blind'' for the complement set $\J^c$. This is because Alice is constrained to using a stealthy codebook, and hence any set of $Z$ links must have marginal distributions that look innocent. For instance, if James controls $2$ out of $5$ links (say links $1$ and $2$), he knows that Alice's transmissions on any other link $j \notin \{1,2\}$ must have joint distribution with links in $\{1,2\}$ according to
the innocent distribution. 

\subsection{Achievability Scheme} \label{sec:scheme}
	The achievability scheme relies on a {\it random coding argument}. Let $P_X$ be the optimal distribution in optimization (B).

\noindent{\it \underline{Encoder:}} We set 
$R = \underline{K}(P^{\text{inn}}_X,Z) - \epsilon = \min_{\J \in \mathfrak{J}} H(X_{\J^c}) - \epsilon$ 
for some sufficiently small $\epsilon > 0$ such that $R > \max_{\J \in \mathfrak{J}} H(X_{\J})$. For each message $m$, the codeword $\x(m)$ is generated according to the $n$-letter product distribution $P_\X \triangleq \prod_{i=1}^n P_X$. Alice encodes $m$ to $\x(m)$, and transmits $\x(m)$ over the multipath network. The codebook $\C$ is a collection of codewords $\{\x(m)\}_{m=1}^N$; for any set $\J$, we denote the codebook subject to the set $\J$ as $\C_{\J} \triangleq \{\x_{\J}(m)\}_{m=1}^N$.

\noindent{\it \underline{Decoder:}} Since Bob does not know the set $\J$ controlled by James {\it a priori}, he attempts to decode based on every possible choice of $\widehat{\J} \in \mathfrak{J}$ and applies an {\it erasure-like decoding} on its corresponding {\it decoding set} $\widehat{\J}^c$. For a specific $\widehat{\J}$, Bob outputs a message $m$ to his list $\mathcal{L}$ if its corresponding sub-codeword $\x_{\widehat{\J}^c}(m)$ on the decoding set $\widehat{\J}^c$ equals $\y_{\widehat{\J}^c}$. 
This procedure is repeated for every $\widehat{\J} \in \mathfrak{J}$. Bob decodes to $\widehat{T}=1$ and $\widehat{M}=m$ if the list $\mathcal{L}$ contains a unique message $m$, decodes to $\widehat{T}=0$ and $\widehat{M}=0$ if the list $\mathcal{L}$ is empty, and declares an error otherwise.

\subsection{Proof of Stealth}

Recall that under erasure jamming, we have shown in Section~\ref{sec:sketch_stealth} that the achievability scheme with codebook generation distribution $P_U$ and artificial noisy channel $P_{X|U}$ at the encoder ensures stealth (i.e., $\lim_{n \to \infty}\V(\widehat{P}_{\X_{\J}}, P^{\text{inn}}_{\X_{\J}}) = 0$ for every $\J \in \mathfrak{J}$) as long as the rate  $R > \max_{\J \in \mathfrak{J}} I(U;X_{\J})$. Note that the above result holds for any $P_U$ and $P_{X|U}$. 

To prove the stealth of the achievability scheme proposed in this section for overwrite jamming, we can simply reuse the result for erasure jamming by replacing $P_U$ with $P_X$ and replacing $P_{X|U}$ with a noiseless channel (i.e., $P_{X|U}(x|u) = \mathbbm{1}\{x = u\}$), thus the stealth is guaranteed as long as the rate $R$ is larger than  $\max_{\J \in \mathfrak{J}} I(X;X_{\J}) = \max_{\J \in \mathfrak{J}} H(X_{\J})$. By noting that we set $R > \max_{\J \in \mathfrak{J}} H(X_{\J})$ in our scheme, the proof of stealth is then completed.

\subsection{Proof of Reliability}
When Alice is active ($T = 1$), we first assume $M = m$ is transmitted and the subset $\J \in \mathfrak{J}$ is controlled by James. We consider the following two cases.

{\bf Case 1:} When Bob decodes according to the ``correct'' decoding set $\widehat{\J}^c = \J^c$, the transmitted message $m \in \mathcal{L}$ since the subset $\J^c$ is noiseless and $\x_{\J^c}(m)$ must equal Bob's observations $\y_{\J^c}$.  In this case, error occurs if there exists a message $m' \ne m$ such that $\x_{\J^c}(m') = \x_{\J^c}(m)$. However, since the rate $R < H(P_{X_{\J^c}})$, it can be shown that the probability of error is vanishing (which can also be viewed as a consequence of the channel coding theorem with an input distribution $P_{X_{\J^c}}$ and a noiseless channel).

{\bf Case 2:}  When Bob decodes according to any  other ``incorrect'' decoding set $\widehat{\J}^c$ ($\widehat{\J}^c \ne \J^c$), we prove that with high probability, no other message $m' \ne m$ falls into $\mathcal{L}$. We make it concrete in the following. For any $\widehat{\J}^c \ne \J^c$, we partition $\widehat{\J}^c$ into disjoint subsets $\G$ and $\B$, where $\G \triangleq \widehat{\J}^c \cap \J^c$ is the ``good set'', while $\B \triangleq \widehat{\J}^c \cap \J$ is the ``bad set''. For simplicity we consider the worst case wherein $\B = \J$ (i.e., the decoding set $\widehat{\J}^c$ contains all the links controlled by James), thus $\widehat{\J}^c = \G \cup \B = \G \cup \J$. James is able to replace $\x_{\J}$ with $\y_{\J}$ according to an arbitrary distribution $W_{\Y_{\J}|\X_{\J},\C}$. We denote Bob's observation on the decoding set $\wjc$ by $\y_{\wjc} \triangleq (\x_{\G}(m), \y_{\J}),$ since the observations on $\G$ corresponds to the sub-codeword $\x_{\G}(m)$ of the transmitted message $m$.
Hence, the average probability of error with respect to $\widehat{\J}^c$ and $W_{\Y_{\J}|\X_{\J},\C}$ is given by   
\begin{align}
\sum_{m=1}^N \frac{1}{N}\sum_{\y_\J \ne \xj }W(\y_{\J}|\x_{\J}(m),\C)\mathbbm{1}\{(\x_{\G}(m),\y_{\J})\in \C_{\wjc} \}, \label{eq:sun}
\end{align}
where the indicator function equals one if Bob's observation $\y_{\wjc}$ lies in $\C_{\wjc}$ (or equivalently, there exists a message $m' \ne m$ such that $\x_{\G}(m') = \x_{\G}(m)$ and $\x_{\J}(m') = \y_{\J}$). Note that we exclude $\y_{\J} = \xj$ in~\eqref{eq:sun}, since no decoding error would  occur if $\y_{\J} = \xj$ (i.e., James does not jam anything). By partitioning all $\x_{\J}$ into typical and atypical sets and gathering all messages with the same sub-codeword $\x_{\J}$ together, one can bound~\eqref{eq:sun} from above as 
\begin{align}
&\sum_{\xj \in \AXj} \sum_{m:\xj(m)=\xj}\frac{1}{N}\sum_{\y_{\J} \ne \xj}W(\y_{\J}|\x_{\J},\C) \mathbbm{1}\{(\x_{\G}(m),\y_{\J})\in \C_{\wjc} \} + \sum_{\xj \notin \AXj} \sum_{m:\xj(m)=\xj}\frac{1}{N} \notag \\
&=\frac{1}{N}\sum_{\xj \in \AXj} \sum_{\y_\J \ne \xj}W(\y_{\J}|\x_{\J},\C) \sum_{m:\xj(m)=\xj}\mathbbm{1}\{(\x_{\G}(m),\y_{\J})\in \C_{\wjc} \} + \frac{|m:\xj(m) \notin \AXj|}{N} \notag \\
&=\frac{1}{N}\!\!\sum_{\xj \in \AXj} \sum_{\y_\J \ne \xj}\! W(\y_{\J}|\x_{\J},\C)  \big|m\!:\! \{\xj(m)=\xj\} \cap \{(\x_{\G}(m),\y_{\J}) \in \C_{\wjc}\} \big| + \frac{|m:\xj(m) \notin \AXj|}{N}. \label{eq:ten}
\end{align}

\begin{lemma}\label{lemma:ratio}
	For any $\y_{\J}$ and typical $\x_{\J} \in \AXj$, with probability $1 - 2^{-\omega(n)}$ (i.e., super-exponentially close to one) over the code design, a randomly chosen code $\C$ satisfies 
	\begin{align}
	\frac{\big|m:\{\xj(m)=\xj\} \cap \{(\x_{\G}(m),\y_{\J}) \in \C_{\wjc}\} \big|}{\big| m:\xj(m)=\xj \big|} \le \varepsilon'_n, \label{eq:face}
	\end{align}
	where $\varepsilon'_n \to 0$ as $n \to \infty$.
\end{lemma}

Lemma~\ref{lemma:ratio} is the crux of our proof, and is formally proved in Appendix~\ref{proof:reliability}. Although showing that on expectation the left-hand side (LHS) of~\eqref{eq:face} is a decaying function of $n$ is relatively straightforward, it is much trickier to prove that the probability that the LHS of~\eqref{eq:face} is a decaying function of $n$ is super-exponentially close to one (which is essential for taking a union bound over exponentially many $\x_J$ and $\y_J$ in the next step). This is because one cannot apply many standard concentration inequalities (such as the Chernoff bound and the Hoeffding's inequality) to the numerator in~\eqref{eq:face} owing to the dependence issue. To circumvent this dependence issue, we first represent the numerator in~\eqref{eq:face} by a function $F(\cdot)$ (to be defined in~\eqref{eq:f} of Appendix~\ref{proof:reliability}) of a subset of codewords, and then apply the McDiarmid's inequality to concentrate the numerator. 

By taking a {\it union bound} over exponentially many $\y_{\J}$ and typical $\xj$, we have that no matter which typical $\xj$ is received and which $\y_{\J}$ is overwritten by James, the induced probability of error is always bounded from above by $\varepsilon'_n$. Therefore, with probability $1-2^{-\omega(n)}$, the first term of~\eqref{eq:ten} can be bounded from above by
\begin{align*}
\frac{\varepsilon'_n}{N}\sum_{\xj \in \AXj} \sum_{\y_\J}W(\y_{\J}|\x_{\J},\C) \cdot \big|m:\xj(m)=\xj \big| \le \varepsilon'_n,
\end{align*}    
for \emph{any} conditional distribution $W_{\Y_{\J}|\X_{\J},\C}$.
It then remains to bound the second term of~\eqref{eq:ten}.
\begin{lemma}\label{lemma:new}
	With probability with probability $1 - 2^{-\omega(n)}$ (i.e., super-exponentially close to one) over the code design, a randomly chosen code $\C$ satisfies $\frac{|m:\xj(m) \notin \AXj|}{N} \le \frac{3}{2}\gamma$, where $\gamma \to 0$  as $n \to \infty.$
\end{lemma}
Lemma~\ref{lemma:new} is proved in Appendix~\ref{appendix:lemma3}.
Based on Lemma~\ref{lemma:new}, one can show that the average probability of error with respect to $\widehat{\J}^c$ and any conditional distribution $W_{\Y_{\J}|\X_{\J},\C}$ is vanishing. Note that we need to consider all possible decoding sets $\widehat{\J}^c \ne J^c$. A union bound over all decoding sets $\widehat{\J}^c \in \mathfrak{J}^c$ yields that with high probability, there does not exist a fake message $m' \ne m$ falling into $\mathcal{L}$, which in turn implies the list $\mathcal{L}$ contains the correct message $m$ only. 

When Alice is innocent ($T=0$), a similar proof technique shows that the list $\mathcal{L}$ is empty with high probability. This completes the proof of reliability.  \qed

\subsection{Discussion}
	It would be interesting to see if it is possible to modify the proof technique above to show that the rate $\bar{K}(P^{\mathrm{inn}}_X,Z)-\epsilon$ is also achievable under overwrite jamming. The main challenge is to deal with the complicated joint typicality relationship among $(\un, \y_{\J},\x_{\G}$), since we introduce an auxiliary variable $U$ and use typicality decoding. We believe that this proof strategy likely works and conjecture the following achievability. 

\begin{conjecture}
	For any $P^{\text{inn}}_X$ and non-negative integer $Z < C/2$, the rate $R = \bar{K}(P^{\emph{inn}}_X,Z)-\epsilon$ is achievable under overwrite jamming for any small $\epsilon >0$.
\end{conjecture}

\section{Conclusion and Future Directions}

This work investigates the problem of stealthy communication over an adversarially jammed multipath network. We first present a coding scheme that is robust to the erasure jamming attack. Subsequently, we show that even when the adversary is able to arbitrarily overwrite the transmissions on links that he controls (i.e., under the overwrite jamming model), perhaps surprisingly, a positive rate is also achievable. For both achievability schemes, we provide rigorous proofs for both stealth and reliability.

Finally, we put forth two promising directions for future work.
\begin{enumerate}
	\item One would expect to verify the correctness of Conjecture 1 by proving that the coding scheme used for erasure jamming is also applicable to the overwrite jamming attack.   
	\item Another direction that is worth exploring is to characterize the stealthy capacities by developing tight information-theoretic upper bounds for this stealthy communication problem. 
\end{enumerate}

\appendices
\section{Chernoff bound}	\label{appendix:chernoff}
Let $Q_1, Q_2, \ldots, Q_n$ be independent (but not necessarily identically distributed) random variables taking values in $\{0,1\}$, and $Q =\sum_{i=1}^n Q_i$. Then, for any $\epsilon \in (0,1)$,
\begin{align*}
&\PP\left(Q \ge (1+\epsilon)\mathbb{E}(Q)\right) \le \exp\left(-\frac{\epsilon^2 \mathbb{E}(Q)}{3}\right), \\
&\PP\left(Q \le (1-\epsilon)\mathbb{E}(Q)\right)\le \exp\left(-\frac{\epsilon^2 \mathbb{E}(Q)}{2}\right) \le \exp\left(-\frac{\epsilon^2 \mathbb{E}(Q)}{3}\right).
\end{align*}

\section{Preliminaries}
\begin{definition}
	The {\it $\gamma$-strongly typical set} $\A^{n,\gamma}_{X}$ with respect to $P_X$ is the set of $\x \in \mathcal{X}^n$ such that $N(x;\x)=0$ if $P_X(x) =0$, and $\sum_{x \in \mathcal{X}} \left|\frac{N(x;\x)}{n}-P_X(x)\right| \le \gamma$, where $N(x;\x)$ is the number of occurrences of $x$ in $\x$, and $\gamma \to 0$ as $n \to \infty$. 
\end{definition}
The $\gamma$-strongly typical sets $\A_{U}^{n,\gamma}$ and $\A_{X_{\J}}^{n,\gamma}$ (with respect to $P_{U}$ and $P_{X_{\J}}$ respectively) are defined in a similar way.  

\begin{definition}
	The {\it $\gamma$-strongly jointly typical set} $\A^{n,\gamma}_{UX}$ with respect to $P_{UX}$ is the set of $(\un,\x) \in \mathcal{U}^n \times \mathcal{X}^n$ such that $N(u,x;\un,\x)=0$ if $P_{UX}(u,x) =0$, and 
	$\sum_{u \in \mathcal{U}}\sum_{x \in \mathcal{X}} \left|\frac{N(u,x;\un,\x)}{n}-P_{UX}(u,x)\right| \le \gamma,$
	where $N(u,x;\un,\x)$ is the number of occurrences of $(u,x)$ in $(\un,\x)$. 
\end{definition}

\begin{definition}
	For any fixed typical $\x$, We say $\un \in \A^{n,\gamma}_{U \x}$ if $(\un,\x) \in \A^{n,\gamma}_{UX}.$
\end{definition}
We define the $\gamma$-strongly typical sets  $\A_{U}^{n,\gamma}, \A_{X_{\J}}^{n,\gamma}, \A_{X_{\G}}^{n,\gamma},$ and $\gamma$-strongly jointly typical set $\A^{n,\gamma}_{UX_\J}$,$\A^{n,\gamma}_{X_{\G}X_\J}$ in a similar way.

\begin{remark}
	It is worth noting that if $(\un,\x) \in \A^{n,\gamma}_{UX}$, then both $\un \in \A_{U}^{n,\gamma}$ and $\x \in \A_{X}^{n,\gamma}$.
\end{remark}

\section{Proof of Stealth for Erasure Jamming} \label{proof:stealth}
Note that the $n$-letter innocent distribution $\pixj$ on the subset $\J$ equals the stochastic processes $P_\U$ and $P_{\X_{\J}|\U}$ simulated by the encoder $\Psi$. For a fixed $\xj$, by considering conditionally typical $\un$ and atypical $\un$, we have 
\begin{align}
\pixj & = \sum_{\un}P_\U(\un)\pxju = \sum_{\un \in \AU}P_\U(\un)\pxju   + \sum_{\un \notin \AU}P_\U(\un)\pxju. \notag 
\end{align}
The active distribution $\phxj$ on the subset $\J$ (induced by the code $\C$) equals
\begin{align}
\phxj &= \sum_{i=1}^N \frac{1}{N} \pxjum  \notag \\
&= \sum_{m: \un(m) \in \AU}\frac{1}{N}\pxjum  + \sum_{m: \un(m) \notin \AU}\frac{1}{N} \pxjum. \notag
\end{align}

Recall that the variational distance between $\pixj$ and $\phxj$ equals
\begin{align}
&\V\left(\pin, \ph \right) = \frac{1}{2} \sum_{\xj \in \AXj}\left|\pixj - \phxj \right| + \frac{1}{2} \sum_{\xj \notin \AXj}\left|\pixj - \phxj \right| \label{eq:cong1} \\
&\le \underbrace{\frac{1}{2} \sum_{\xj \in \AXj}\left|\pixj - \phxj \right|}_{\text{Term } (C)}  + \underbrace{\frac{1}{2} \sum_{\xj \notin \AXj}\pixj}_{\text{Term } (D)} + \underbrace{\frac{1}{2} \sum_{\xj \notin \AXj}\phxj}_{\text{Term } (E)}, \label{eq:cong2}
\end{align}
where~\eqref{eq:cong1} is obtained by dividing $\xj$ into typical $\xj$ and atypical $\xj$, and~\eqref{eq:cong2} follows from the triangle inequality. Note that term $(C)$ can further be bounded from above as 
\begin{align}
(C) &\le \underbrace{\frac{1}{2}\sum_{\xj \in \AXj}\left|\sum_{\un \in \AU}P_\U(\un)\pxju -\sum_{m: \un(m) \in \AU}\frac{1}{N}\pxjum\right|}_{\text{Term } (C_1)} \notag \\
&+ \underbrace{\frac{1}{2}\sum_{\xj \in \AXj}\sum_{\un \notin \AU}P_\U(\un)\pxju}_{\text{Term } (C_2)} + \underbrace{\frac{1}{2}\sum_{\xj \in \AXj}\sum_{m: \un(m) \notin \AU}\frac{1}{N}\pxjum}_{\text{Term } (C_3)}. \label{eq:haha}
\end{align}
Term $(D)$ and term $(C_2)$ correspond to $\Xj \notin \AXj$ and $\U \notin \AU$ (for a typical $\xj$), respectively, hence both of the two terms goes to zero as $n$ tends to infinity (by the {\it law of large number}). Term $(E)$ and term $(C_3)$ correspond to similar atypical events but depends on the specific codebook $\C$. Prior work~\cite{CheBJ:13} showed that with high probability over the code design, both of the two terms approach zero as $n$ tends to infinity. We now focus on term $(C_1)$ in the following.
\begin{align}
(C_1) &= \frac{1}{2}\sum_{\xj \in \AXj}\Bigg|\sum_{\un \in \AU}P_\U(\un)\pxju -\sum_{m: \un(m) \in \AU}\frac{1}{N}\pxjum \bigg| \notag \\
&= \frac{1}{2}\sum_{\xj \in \AXj}\Bigg|\sum_{\TU} \sum_{\un \in \TU}P_\U(\un)\pxju - \sum_{\TU}\sum_{m: \un(m) \in \TU}\frac{1}{N}\pxjum \Bigg| \notag \\
&= \frac{1}{2}\sum_{\xj \in \AXj}\Bigg|\sum_{\TU}\pxju  \cdot \left(P_\U\left(\U \in \TU\right) - \frac{|m: \un(m)\in \TU|}{N}\right) \Bigg|. \label{eq:chun}
\end{align}
Due to the linearity of expectation, we have
\begin{align}
\mu \triangleq \E_{\C}\left(|m: \un(m)\in \TU|\right) = N\cdot P_\U\left(\U \in \TU\right), 
\end{align}
which is exponentially large since 
$N=2^{nR} > 2^{nI(U;X_{\J})}$ and 
$P_\U\left(\U \in \TU\right) \stackrel{\cdot}{=} 2^{nH(U|X_\J)}/2^{nH(U)} = 2^{-nI(U;X_{\J})}$.
Since the codewords $\un(m)$ are chosen independently, by the Chernoff bound we have
\begin{align*}
\PP_{\C}\left(\left|\frac{|m: \un(m)\in \TU|}{\mu}-1\right| \le \varepsilon_1(n) \right) \ge 1-2e^{-\frac{1}{3}\mu \varepsilon_1^2(n)},
\end{align*} 
where $\varepsilon_1(n) \to 0$ as $n \to \infty$. For instance, we set $\varepsilon_1(n) = n^{-1}$. Hence 
\begin{align}
&\PP_{\C}\Bigg(\left|P_\U\left(\U \in \TU\right) - \frac{|m: \un(m)\in \TU|}{N}\right|  \le n^{-1} P_\U\left(\U \in \TU\right) \Bigg) \ge 1-2e^{-\frac{\mu}{3n^2} }. \label{eq:ad}
\end{align}
Replacing~\eqref{eq:ad} into~\eqref{eq:chun}, we have
\begin{align}
(C_1)  \stackrel{\text{w.h.p.}}{=} \frac{1}{2n}\sum_{\xj \in \AXj}\left|\sum_{\TU}\pxju P_\U\left(\U \in \TU\right) \right| &= \frac{1}{2n}\sum_{\xj \in \AXj}\sum_{\TU}\sum_{\un \in \TU} \pxju P_{\U}(\un) \notag\\
&\le \frac{1}{2n} \sum_{\xj}\sum_{\un}\pxju P_{\U}(\un) \le \frac{1}{2n}. \notag
\end{align}

By combining $(C_1), (C_2), (C_3), (D), (E)$, we eventually show that with high probability over the code design, a randomly chosen code $\C$ satisfies $\lim_{n \to \infty} \V(P^{\text{inn}}_{\X_{\J}}, \widehat{P}_{\X_{\J}}) = 0$ for every $\J \in \mathfrak{J}$.

\section{Proof of Lemma~\ref{lemma:ratio}} \label{proof:reliability}

By the {\it strong asymptotic equipartition property} (strong AEP), we know that for any typical $\xj$, there exists $\eta_{\gamma} > 0$ such that $\eta_{\gamma} \to 0$ as $\gamma \to 0$ and
\begin{align*}
2^{-n(H(X_{\J})+\eta_{\gamma})} \le P_{\X_{\J}}(\X_{\J}=\xj) \le 2^{-n(H(X_{\J})-\eta_{\gamma})}.
\end{align*} 
Since $P_X$ satisfies~\eqref{eq:con2} in optimization (B), we have 
$\min_{\J \in \mathfrak{J}} H(X_{\J^c}) > \max_{\J \in \mathfrak{J}} H(X_{\J}) \ge H(X_{\J}).$
Hence, there exists a $\delta > 0$ such that 
$\delta \triangleq \min_{\J \in \mathfrak{J}} H(X_{\J^c}) - H(X_{\J}).$
We let $\epsilon \ll \delta$ and $\eta_{\gamma} \ll \delta$.

\begin{claim} \label{claim:1}
	For any typical $\xj$, with probability $1-2^{-\omega(n)}$ over the code design, 
	\begin{align*}
	|m:\xj(m) = \xj| \ge (1-n^{-1})\cdot 2^{n(\delta - \epsilon -\eta_{\gamma})}.
	\end{align*}
\end{claim}

\noindent{\it Proof:} The expected number of codewords such that their sub-codeword on $\J$ equals $\xj$ is
\begin{align}
\E_{\C}\left(|m:\xj(m) = \xj| \right) = 2^{nR}\cdot  P_{\X_{\J}}(\X_{\J}=\xj) &\ge 2^{n(\min_{\J \in \mathfrak{J}} H(X_{\J^c})-\epsilon)}\cdot 2^{-n(H(X_{\J})+\eta_{\gamma})} =2^{n(\delta - \epsilon - \eta_{\gamma})}, \notag
\end{align}
which is exponentially large since $\epsilon \ll \delta$ and $\eta_{\gamma} \ll \delta$. Note that each of the codeword is chosen independently, hence by the Chernoff bound,
\begin{align}
&\PP\left(|m:\xj(m) = \xj| \ge (1-n^{-1})2^{n(\delta-\epsilon-\eta_{\gamma})} \right) \notag \\
&\ge \PP\left(|m:\xj(m) = \xj| \ge (1-n^{-1})\E_{\C}(|m:\xj(m) = \xj|) \right) \notag \\
&\ge 1 - \exp\left(-\frac{1}{3}n^{-2}2^{n(\delta-\epsilon-\eta_{\gamma})}\right) = 1 - 2^{-\omega(n)}.\label{eq:shuo} 
\end{align}
\qed

For notational convenience let $\xi \triangleq  \delta-2\epsilon-2\eta_{\gamma}+5\nu_{\gamma}$.
\begin{claim} \label{claim:2}
	For any $\y_{\J}$ and typical $\xj$, with probability $1-2^{-\omega(n)}$ over the code design, 
	\begin{align}
	&\big|m:\{\xj(m) = \xj\} \cap \{(\x_{\G}(m),\y_{\J}) \in \C_{\wjc}\}\big|  \le (1+n^{-1})^3 \cdot 2^{n\xi}.  \notag
	\end{align}
\end{claim}

\noindent{\it Proof:} Let $\mathcal{S} = \big\{m: \{\xj(m) = \xj\} \cup \{\xj(m) = \y_{\J}\} \big\}$ be a subset of messages such that each $m \in \mathcal{S}$ satisfies either $\xj(m)= \xj$ or $\xj(m) = \y_{\J}$. Similar to~\eqref{eq:shuo}, we have 
\begin{align*}
\PP\left(|\mathcal{S}| > 2(1+n^{-1}) 2^{n(\delta-\epsilon-\eta_{\gamma})}\right) \le 2^{-\omega(n)}.
\end{align*}
\newcommand{\EE}{\mathcal{E}}
We denote the events $\{(\x_{\G},\xj) \in \C_{\wjc} \}$ and $\{(\x_{\G},\y_{\J}) \in \C_{\wjc} \}$ by $\EE_{\x_{\G},\xj}$ and $\EE_{\x_{\G},\y_{\J}}$, respectively, and it is worth noting that 
\begin{align}
&\big|m:\{\xj(m) = \xj\} \cap \{(\x_{\G}(m),\y_{\J}) \in \C_{\wjc}\} \big|   = \sum_{\x_{\G}} \mathbbm{1} \big\{\EE_{\x_{\G},\xj} \cap \EE_{\x_{\G},\y_{\J}} \big\}.  \notag
\end{align}
Let $\kappa \triangleq \left(1+n^{-1}\right)^3 2^{n\xi}$, we then have
\begin{align}
&\PP\left(\big|m:\{\xj(m) = \xj\} \cap \{(\x_{\G}(m),\y_{\J}) \in \C_{\wjc}\} \big| > \kappa \right) \notag \\
&=\sum_{i=0}^N \PP(|\mathcal{S}| = i) \PP\left(\sum_{\x_{\G}} \mathbbm{1} \big\{\EE_{\x_{\G},\xj} \cap \EE_{\x_{\G},\y_{\J}} \big\} > \kappa \Big| |\mathcal{S}|=i \right) \notag \\
& \le  2^{-\omega(n)} +  \sum_{i=0}^{2(1+\frac{1}{n})2^{n(\delta-\epsilon-\eta_{\gamma})}} \PP(|\mathcal{S}| = i)  \cdot \PP\left(\sum_{\x_{\G}} \mathbbm{1} \big\{\EE_{\x_{\G},\xj} \cap \EE_{\x_{\G},\y_{\J}} \big\} > \kappa \Big| |\mathcal{S}|=i \right).  \label{eq:ha2}
\end{align}
When $|\mathcal{S}| = i$, by symmetry we assume that the event 
$$\Delta_i \triangleq  \{m_1,m_2,\ldots,m_i \in \mathcal{S}, \text{ and } m_{i=1},\ldots, m_{N} \notin \mathcal{S}\}$$ 
occurs. Hence, 
\begin{align}
\PP\left(\sum_{\x_{\G}} \mathbbm{1} \big\{\EE_{\x_{\G},\xj} \cap \EE_{\x_{\G},\y_{\J}} \big\} > \kappa \Big| |\mathcal{S}|=i \right) = \PP\left(\sum_{\x_{\G}} \mathbbm{1} \big\{\EE_{\x_{\G},\xj} \cap \EE_{\x_{\G},\y_{\J}} \big\} > \kappa \Big| \Delta_i \right).
\end{align}
Let's first consider the expectation 
\begin{align}
\E_{\C}\left(\sum_{\x_{\G}} \mathbbm{1} \big\{\EE_{\x_{\G},\xj} \cap \EE_{\x_{\G},\y_{\J}} \big\} \Big| \Delta_i \right) &= \sum_{\x_{\G}} \PP\left(\EE_{\x_{\G},\xj} \cap \EE_{\x_{\G},\y_{\J}} \Big| \Delta_i \right) \notag\\
& \stackrel{n \to \infty}{=} \sum_{\x_{\G} \in \A^{n,\gamma}_{X_{\G}\xj}} \PP\left(\EE_{\x_{\G},\xj} \cap \EE_{\x_{\G},\y_{\J}} \Big| \Delta_i \right) \label{eq:deng} \\
&\le \sum_{\x_{\G} \in \A^{n,\gamma}_{X_{\G}\xj}} \PP\left(\EE_{\x_{\G},\xj} \Big| \Delta_i \right) \PP\left(\EE_{\x_{\G},\y_{\J}} \Big| \Delta_i \right). \label{eq:alin}
\end{align}
Equation~\eqref{eq:deng} follows from the negligibility of conditionally atypical $\x_{\G}$, and inequality~\eqref{eq:alin} is due to the fact that if one codeword is fixed and not equals $(\x_{\G},\y_{\J})$, the probability that the codebook contains $(\x_{\G},\y_{\J})$ decreases. Note that conditioned on $\Delta_i$, for $\x_{\G} \in \A^{n,\gamma}_{X_{\G}\xj}$ and each message $m_j$ (for $j \in \{1,2,\ldots, i\}$), there exists a $\nu_{\gamma} >0$ such that $\nu_{\gamma} \to 0$ as $\gamma \to 0$, and the probability that its corresponding sub-codewords $\X_{\G}(m_j)$ and $\Xj(m_j)$ respectively equal $\x_{\G}$ and $\xj$ is bounded from above as 
\begin{align*}
&\PP\left(\{\X_{\G}(m_j) = \x_{\G}\} \cap \{ \Xj(m_j) =\xj \} | \Delta_i \right) \\
&=\PP\left(\Xj(m_j) =\xj | \Delta_i \right) \cdot \PP\left(\X_{\G}(m_j) = \x_{\G}|\Xj(m_j) =\xj, \Delta_i\right) \le 2^{-n(H(X_{\G}|X_{\J})-\nu_{\gamma})}.
\end{align*}
For $i \in\{1, 2, \ldots, 2(1+n^{-1})2^{n(\delta-\epsilon-\eta_{\gamma})}\}$, we have
\begin{align}
&\PP_{\C}\left( \EE_{\x_{\G},\xj} \Big| \Delta_i \right) \le 1- \left(1- 2^{-n(H(X_{\G}|X_{\J})-\nu_{\gamma})}\right)^i =(1+n^{-1})2^{n(\delta-\epsilon-\eta_{\gamma}-H(X_{\G}|X_{\J})+2\nu_{\gamma})}, \label{eq:food1} \\
&\PP_{\C}\left( \EE_{\x_{\G},\y_{\J}} \Big| \Delta_i \right) \le (1+n^{-1})2^{n(\delta-\epsilon-\eta_{\gamma}-H(X_{\G}|X_{\J})+2\nu_{\gamma})}. \label{eq:food2}
\end{align}
Combining~\eqref{eq:alin},~\eqref{eq:food1}, and~\eqref{eq:food2}, we have
\begin{align}
\E_{\C}\left(\sum_{\x_{\G}}\mathbbm{1} \{ \EE_{\x_{\G},\xj} \cap \EE_{\x_{\G},\y_{\J}} \} \Big| \Delta_i \right) &\le \sum_{\x_{\G} \in \A^{n,\gamma}_{X_{\G}\xj}} (1+n^{-1})^2 \cdot  2^{2n(\delta-\epsilon-\eta_{\gamma}-H(X_{\G}|X_{\J})+2\nu_{\gamma})} \notag \\
&\le (1+n^{-1})^2 \cdot 2^{n\xi}, \label{eq:what}
\end{align}
where~\eqref{eq:what} is obtained by noting $|\A^{n,\gamma}_{X_{\G}\xj}| \le 2^{n(H(X_{\G}|X_{\J})+\nu_{\gamma})}$ and 
\begin{align*}
H(X_{\G}|X_{\J}) = H(X_{\G},X_{\J}) - H(X_{\J}) \ge \min_{\J \in \mathfrak{J}} H(X_{\J^c}) - H(X_{\J}) = \delta.
\end{align*}
We now use the McDiarmid's inequality to concentrate $\sum_{\x_{\G}}\mathbbm{1} \big\{ \EE_{\x_{\G},\xj} \cap \EE_{\x_{\G},\y_{\J}} \big\}$ conditioned on $\Delta_i$.

\begin{lemma}[McDiarmid's inequality~\cite{mcdiarmid1989method}]
	Let $X_1, \ldots, X_n$ be independent random variables taking values in ranges $R_1, \ldots, R_n$, and let $F: R_1 \times \cdots \times R_n \to \mathbb{R}$ be a function with the property that if one freezes all but the $i$-th coordinate of $F(x_1,\ldots,x_n)$ for some $1 \le i \le n$, then $F$ only fluctuates by most $c_i > 0$, i.e., $\big|F(x_1,\ldots,x_{i-1},x_i,x_{i+1},\ldots,x_n) -F(x_1,\ldots,x_{i-1},x'_i,x_{i+1},\ldots,x_n)\big| \le c_i,$
	then for any $\lambda > 0$, one has 
	\begin{align*}
	&\PP(|F(X_1,\ldots,X_n)-\E F(X_1,\ldots,X_n)| \ge \lambda \sigma) \le K \exp(-k \lambda^2),
	\end{align*}
	for some constants $K,k >0$, where $\sigma^2 = \sum_{i=1}^n c_i^2$.
\end{lemma}

Let $\{\X_j\}_{j=1}^i \stackrel{\text{i.i.d.}}{\sim} P_{\X|\X \in \mathcal{S}}$ be the independent random variables corresponding to $\{m_j\}_{j=1}^i$, where $P_{\X|\X \in \mathcal{S}}(\x) = \frac{P_{\X}(\x) \mathbbm{1}\{\x \in \mathcal{S}\}}{P_{\X}(\X \in \mathcal{S})}.$
Let 
\begin{align}
	F(\X_1, \ldots, \X_i) \triangleq \sum_{\x_{\G}}\mathbbm{1} \{ \EE_{\x_{\G},\xj} \cap \EE_{\x_{\G},\y_{\J}} \}. \label{eq:f}
\end{align} 
Note that $\E F(\X_1, \ldots, \X_i) \le (1+n^{-1})^2 \cdot 2^{n\xi}$ by~\eqref{eq:what}, and 
$c_j = 1$ for all $j \in \{1,2,\ldots,i\},$ 
since changing one codeword $\X_j$ can only fluctuate the function $F(\X_1, \ldots, \X_i)$ at most by one. By letting $\lambda = \frac{(1+n^{-1})^{3/2}}{\sqrt{2}n}2^{n(\frac{1}{2}\delta - \frac{3}{2}\epsilon - \frac{3}{2}\eta_{\gamma}+5\nu_{\gamma})}$, we have
\begin{align}
&\PP\left(F(\X_1, \ldots, \X_i) \ge (1+n^{-1}) \E F(\X_1, \ldots, \X_i)\right) \le K \exp\left(-k \lambda^2 \right) = 2^{-\omega(n)}. \notag 
\end{align}
Therefore, we obtain
\begin{align}
\PP_{\C}\Big(\sum_{\x_{\G}}\mathbbm{1} \big\{\EE_{\x_{\G},\xj} \cap \EE_{\x_{\G},\y_{\J}} \big\}  > (1+n^{-1})^3 \cdot 2^{n\xi}\Big| |\mathcal{S}|=i \Big) \le 2^{-\omega(n)}. \label{eq:ha}
\end{align} 
Substituting~\eqref{eq:ha} into~\eqref{eq:ha2} and taking a union bound over all typical size of $|\mathcal{S}|$, we have
\begin{align}
&\PP_{\C}\Big(\big|m:\{\xj(m) = \xj\} \cap \{(\x_{\G}(m),\y_{\J}) \in \C_{\wjc}\} \big|  > (1+n^{-1})^3 \cdot 2^{n\xi} \Big) \le 2^{-\omega(n)}. \notag
\end{align} \qed

Finally, by combining Claims~\ref{claim:1} and~\ref{claim:2} and setting $\eta_{\gamma},\nu_{\gamma} \ll \epsilon$, we have that with probability at least  $1-2^{-\omega(n)}$ over the code design,,
\begin{align*}
&\frac{\big|m:\{\xj(m)=\xj\} \cap \{(\x_{\G}(m),\y_{\J}) \in \C_{\wjc}\} \big|}{\big| m:\xj(m)=\xj \big|}  \le (1+n^{-1})^2\cdot 2^{-n(\epsilon+\eta_{\gamma}-\nu_{\gamma})} = \varepsilon'_n,
\end{align*}
which completes the proof of Lemma~\ref{lemma:ratio}.

\section{Proof of Lemma~\ref{lemma:new}} \label{appendix:lemma3}
First note that $\PP\left(\X_J \notin \AXj \right) < \gamma$ according to the property of the $\gamma$-strongly typical set $\AXj$. Thus, the expected number of codewords that do not belong to the typical set $\AXj$ is 
\begin{align*}
	\E_{\C}\left(|m:\xj(m) \notin \AXj|\right) = N\cdot \PP\left(\X_J \notin \AXj \right) = \gamma N,
\end{align*}
which is exponentially large since $N = 2^{nR}$ and $\gamma = \Omega(1/n)$ by definition.
By applying the Chernoff bound, we have that 
\begin{align*}
	\PP\left(|m:\xj(m) \notin \AXj| \le \frac{3}{2} \gamma N \right) \ge 1 - \exp\left(-\frac{\gamma N}{12}\right),
\end{align*}
which completes the proof of Lemma~\ref{lemma:new}.

\section{Cardinality bound} \label{appendix:cardinality}

\newcommand{\pscalXcondU}[1]{P_{X|U}\left( #1 \right)} 
\newcommand{\pscalXcondUpri}[1]{P_{X|U'}\left( #1 \right)} 
\newcommand{\pscalU}[1]{P_{U}\left( #1 \right)} 
\newcommand{\pscalUpri}[1]{P_{U'}\left( #1 \right)} 
\newcommand{\pscalXJ}[1]{P_{X_{\J}}\left( #1 \right)} 
\newcommand{\pscalXJc}[1]{P_{X_{\J^c}}\left( #1 \right)} 

\newcommand{\bigJ}{\mathfrak{J}}
\newcommand{\calU}{\mathcal{U}}
\newcommand{\calX}{\mathcal{X}}
\newcommand{\XJ}{X_{\J}}
\newcommand{\XJc}{X_{\J^c}}
\newcommand{\xJ}{x_{\J}}
\newcommand{\xJc}{x_{\J^c}}

This appendix shows that the cardinality of the auxiliary random variable $U$ in optimization (A) is finite. The proof relies on the {\it support lemma}~\cite{el2011network}. Consider any $(U,X)$ defined over $\calU \times \calX$ that satisfies the constraints in optimization (A), where $\calU$ can be arbitrary and the probability density function of $U$ is denoted by $F_U$. Let $\{P_{X|U=u}\}_{u\in \calU} \in \mathcal{P}(\calX)$ be a collection of conditional PMFs on $\calX$. For $\pi \in \mathcal{P}(\calX)$, we have the following $|\calX| + 2|\bigJ| -1$ continuous functions 
\begin{align*}
&\left\{g^{(1)}_j(\pi)\right\}_{j \in \{1,\cdots,|\calX| - 1\}} \triangleq \{\pi(j)\}_{j \in \{1,\cdots,|\calX| - 1\}},  \\
&\left\{g^{(2)}(\pi)\right\}_{J \in \mathfrak{J}} \triangleq \{H(\pi(X_J))\}_{J \in \mathfrak{J}},\quad \left\{g^{(3)}(\pi)\right\}_{J \in \mathfrak{J}} \triangleq \{H(\pi(X_J^c))\}_{J \in \mathfrak{J}},
\end{align*}
where $\pi(X_J)$ and $\pi(X_J^c)$ respectively denote the marginal distributions of $\pi$ on set $J$ and set $J^c$.
Note that the first group of functions are continuous, and the last two groups of functions are also continuous in $\pi$ due to the continuity of entropy function.

By the support lemma, there exists a random variable $U'$ with distribution $P_{U'}$ satisfying $|\mathcal{U}'| = |\calX| + 2|\bigJ| - 1$, and 
\begin{align}       
&P_{X}\left( x \right) = \int_\calU \pscalXcondU{x|u}dF_U(u) = \sum_{u' \in \mathcal{U}'} \pscalXcondUpri{x|u'}\pscalUpri{u'}, \ \forall x \in \calX, \label{eq:px_equal} \\
&H({\XJ|U}) = \int_\calU H({\XJ|U=u})dF_U(u)  = \sum_{u' \in \mathcal{U}'} H({\XJ|U'=u'})\pscalUpri{u'} =H({\XJ|U'}), \ \forall \J \in \bigJ, \notag \\
&H({\XJc|U}) = \int_\calU H({\XJc|U=u})dF_U(u)  = \sum_{u' \in \mathcal{U}'} H({\XJc|U'=u'})\pscalUpri{u'} =H({\XJc|U'}), \ \forall \J \in \bigJ. \notag
\end{align}  
From~\eqref{eq:px_equal}, we note that $H(X)$, $H({\XJ})$, and $H({\XJc})$ are preserved, and we then have
\begin{align*}  
I(\XJ;U) = I(\XJ;U'), \ \forall \J \in \bigJ; \quad \text{and} \quad I(\XJc;U) = I(\XJc;U'), \ \forall \J \in \bigJ.
\end{align*}
Therefore, the random variable pair $(U',X)$ also satisfies the constraints in optimization (A). 

\ifCLASSOPTIONcaptionsoff
  \newpage
\fi



%
%

\bibliographystyle{IEEEtran}
\bibliography{definitions}

%




\end{document}